\documentclass[sn-mathphys,Numbered]{sn-jnl}

\usepackage{bm}
\usepackage{graphicx}%
\usepackage{multirow}%
\usepackage{amsmath,amssymb,amsfonts}%
\usepackage{amsthm}%
\usepackage{mathrsfs}%
\usepackage[title]{appendix}%
\usepackage{xcolor}%
\usepackage{textcomp}%
\usepackage{manyfoot}%
\usepackage{booktabs}%
\usepackage{algorithm}%
\usepackage{algorithmicx}%
\usepackage{algpseudocode}%
\usepackage{listings}%
\usepackage{ulem}
\usepackage{caption}
\usepackage{graphicx}
\usepackage{subfigure}
\usepackage[export]{adjustbox}
\usepackage{ulem}
\usepackage{pdfpages}

\theoremstyle{thmstyleone}%
\theoremstyle{thmstyletwo}%

\theoremstyle{thmstylethree}%

\begin{document}

\title[Article Title]{Controlling surface acoustic waves (SAWs) via temporally graded metasurfaces}



%

%

%
%




\author[1]{\fnm{Jonatha} \sur{Santini}}\email{jonatha.santini@polimi.it}

\author[2]{\fnm{Xingbo} \sur{Pu}}\email{xingbo.pu2@unibo.it}

\author[2]{\fnm{Antonio} \sur{Palermo}}\email{antonio.palermo6@unibo.it}

\author[1]{\fnm{Francesco} \sur{Braghin}}\email{francesco.braghin@polimi.it}

\author*[1]{\fnm{Emanuele} \sur{Riva}}\email{emanuele.riva@polimi.it}

\affil*[1]{\orgdiv{Department of Mechanical Engineering}, \orgname{Politecnico di Milano}, \orgaddress{\street{Via La Masa,1}, \city{Milano}, \postcode{20156}, \country{Italy}}}

\affil[2]{\orgdiv{Department of Civil, Chemical, Environmental and Materials Engineering}, \orgname{University of Bologna}, \orgaddress{\city{Bologna}, \postcode{40136}, \country{Italy}}}


\abstract{In this manuscript, the temporal rainbow effect for surface acoustic waves (SAW) is herein illustrated through a temporal analog of space metagradings. We show that a time-modulated array of mechanical resonators induces a wavenumber-preserving frequency transformation which, in turn, dictates Rayleigh-to-Shear wave conversion.
The process is unfolded through the adiabatic theorem, which allows us to delineate the transition between a solely frequency-converted wave packet and a temporally-driven mode conversion. 
In other words, our paper explores the role of time modulation in the context of elastic metasurfaces, and we envision our implementation to be suitable for designing a new family of SAW devices with frequency conversion, mode conversion, and unusual transport capabilities.}




\maketitle

\section{Introduction}\label{sec1}



Waves, whether they take the form of sound, light, or mechanical vibrations, serve as fundamental carriers of information and energy, thus enabling a plethora of applications across several disciplines. Not surprisingly, the opportunity to manipulate wave motion has had far-reaching implications in diverse technologies that we experience every day for communication, sensing, and isolation purposes \cite{oudich2023tailoring}. Within this framework, metamaterials have blossomed a groundbreaking advancement in material engineering, whereby intricate structural designs have promoted unconventional properties and ensuing wave propagation functions, propelling the quest for new underpinning behaviors. See, for example, the recent advances in the context of topological phases of matter, waveguiding, and nonreciprocal materials \cite{oudich2023tailoring,nassar2020nonreciprocity}. 

Among the available physical platforms, we herein operate with surface acoustic waves (SAWs), which are known to be particularly attractive for signal-processing applications \cite{campbell1991surface}. Indeed, SAW devices are frequently integrated with solid-state platforms as filters, sensors, and delay lines \cite{shao2020non,ferreira2009acoustic,weigel2002microwave} and, in addition, have been conceived to operate with quantum systems such as superconducting circuits, optical components, and optomechanical devices \cite{manenti2017circuit,forsch2020microwave,navarro2022room,satzinger2018quantum}.

An invaluable tool for manipulating surface waves consists of mechanical metasurfaces, which are made of a host medium equipped with periodic tessellations of units placed atop an otherwise homogeneous solid, including features such as holes, rigid cylinders, spheres, or resonant arrays \cite{palermo2016engineered,zeighami2021rayleigh}. Non-periodic configurations are also employed to provide a diverse degree of controllability, whereby the spatial distribution of a relevant parameter is key in wave manipulation. The concept, known as \textit{space metagrading}, has been applied in several frameworks, including studies on 1D structures with rainbow trapping capabilities \cite{santini2022harnessing,alshaqaq2020graded}, which is a phenomenon enabling vibration confinement as a result of the gradual variation of wave velocity in space. Such variation dictates a frequency-preserving wavenumber transformation and is responsible for the formation of ``slow waves" \cite{de2021selective,riva2023adiabatic}. 
Spatially varying metasurfaces have been explored also in the realm of SAWs where the rainbow trapping and underlying wavenumber transformation are accompanied by mode conversion between a wave sitting on the surface and a bulk wave propagating toward the interior of the material \cite{colombi2016seismic,colombi2017enhanced,alan2019programmable}.

Another promising way to manipulate wave motion leverages the temporal degree of freedom, where the modulation does not occur in space, but over time instead. Time-modulation schemes are known to produce frequency conversion, unidirectional waves \cite{marconi2020experimental,palermo2020surface,attarzadeh2020experimental,wu2021non}, temporal compression \cite{riva2023adiabatic2} and pumping \cite{riva2021adiabatic,xia2021experimental}, which are behaviors more easily observable in 1D media.  
2D media with time-varying properties are more intriguing and challenging but, at the same time, allow waveguiding in a controllable manner. A notable example, recently explored both in phononics \cite{santini2023elastic} and photonics \cite{pacheco2020temporal}, consists of tailored time-modulation schemes able to modify the poynting vector which, in turn, promotes wave steering. Other forms of time-modulation functions include periodic drives for mode conversion of light \cite{galiffi2020wood}. 

Motivated by the intriguing waveguiding capabilities of time-varying materials and by the recent discoveries in the context of space and time metagradings, we herein merge these two concepts to produce a solid for frequency and mode conversion. We show that a time-dependent array of mechanical resonators placed atop an elastic half-space can be used to provide different degrees of controllability of wave motion, which depends on the modulation scheme. The dynamics of such a system is herein studied through a temporal analog of space metagradings which, in contrast to its spatial counterpart, dictates a wavenumber-preserving frequency conversion during wave motion.
The underlying frequency transformation is herein unfolded by the adiabatic theorem \cite{santini2023elastic,xia2021experimental,riva2021adiabatic,riva2023adiabatic,bosch2023differences}, delineating the limiting modulation speed for the transformation to occur without energy scattering, which would otherwise imply leakage toward the other wave modes populating the solid. 

In this paper, we show how a time-modulated metasurface can be tailored to produce: (i) pure frequency conversion; (ii) frequency conversion with scattering toward other surface modes; (iii) frequency conversion accompanied by surface-to-bulk mode conversion; (iv) energy transport between two metasurfaces via adiabatic passage. All these aspects will be studied in light of the adiabatic theorem, which provides a suitable mathematical tool for the interpretation of the underlying dynamics. 

Our work provides guidelines for designing time-varying metasurfaces with novel functionalities, such as frequency conversion, mode conversion, and unusual transport properties, and may find new exciting applications in the framework of SAW technologies.

The paper is organized as follows. In section 2, we delineate the transition between adiabatic (slow) and nonadiabtic (fast) transformations for frequency conversion. 
In section 3, we study the interplay between frequency and surface-to-bulk mode conversion, with emphasis on the role of the modulation velocity. Concluding remarks are presented in section 4. 

\section{Frequency conversion induced by slow and fast time-modulation}
\label{sec2}

\begin{figure}
    \hspace{-.0\textwidth}
    \includegraphics[width=1\textwidth]{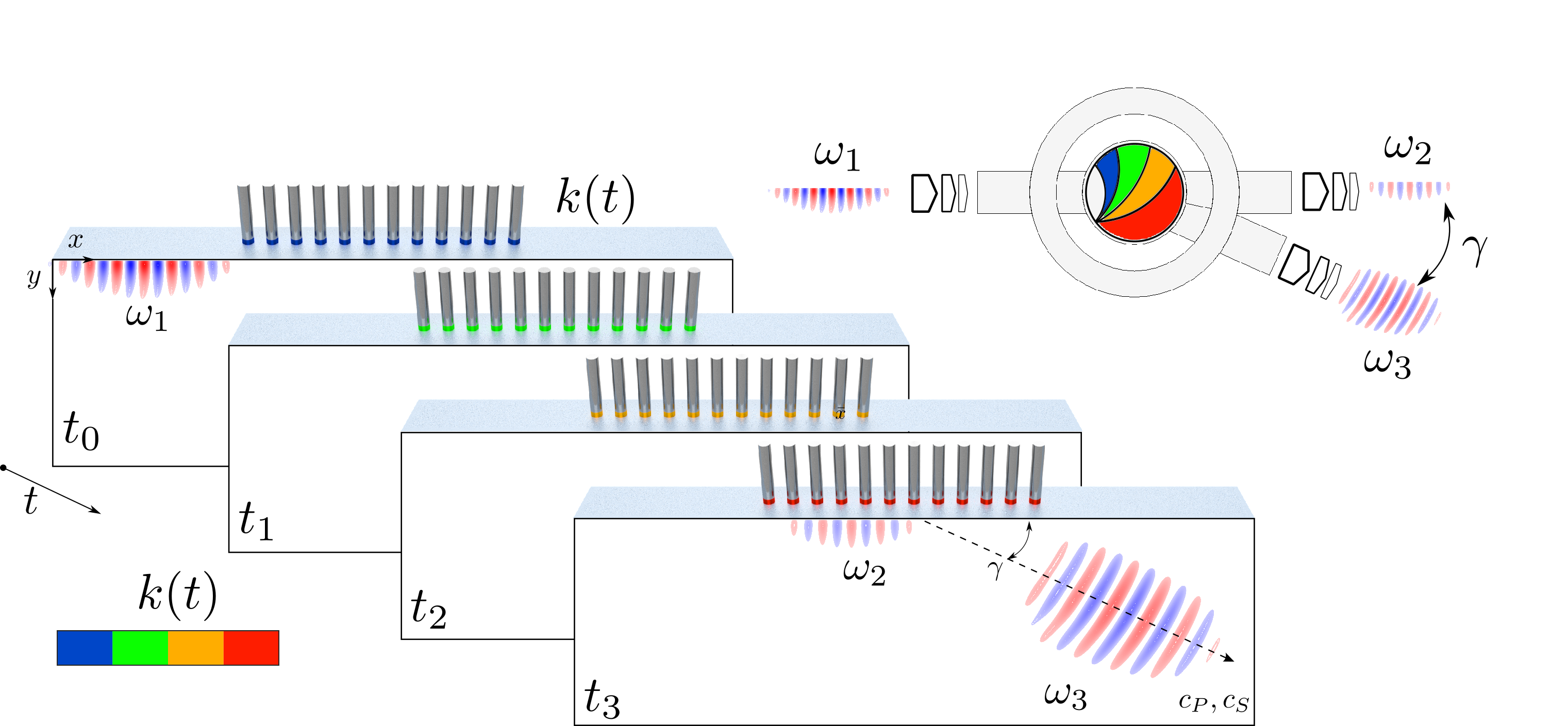}
    \vspace{0.1cm}
    \caption{Schematic of the system. A time-varying metasurface is realized by way of time-dependent mechanical resonators rigidly connected atop an elastic solid which, in a high-frequency regime, emulates a half-space. Time modulation is employed to frequency-convert and mode-convert an impinging surface wave into a bulk wave that propagates towards the bulk of the material with an inclination angle $\gamma$ dictated by the modulation parameters.}
    \label{fig: concept}
\end{figure}
\begin{figure}
    \centering
    \begin{minipage}{1\textwidth}
        \subfigure[\label{fig: adiabatic case dispersion}]{\includegraphics[width=.33\textwidth]{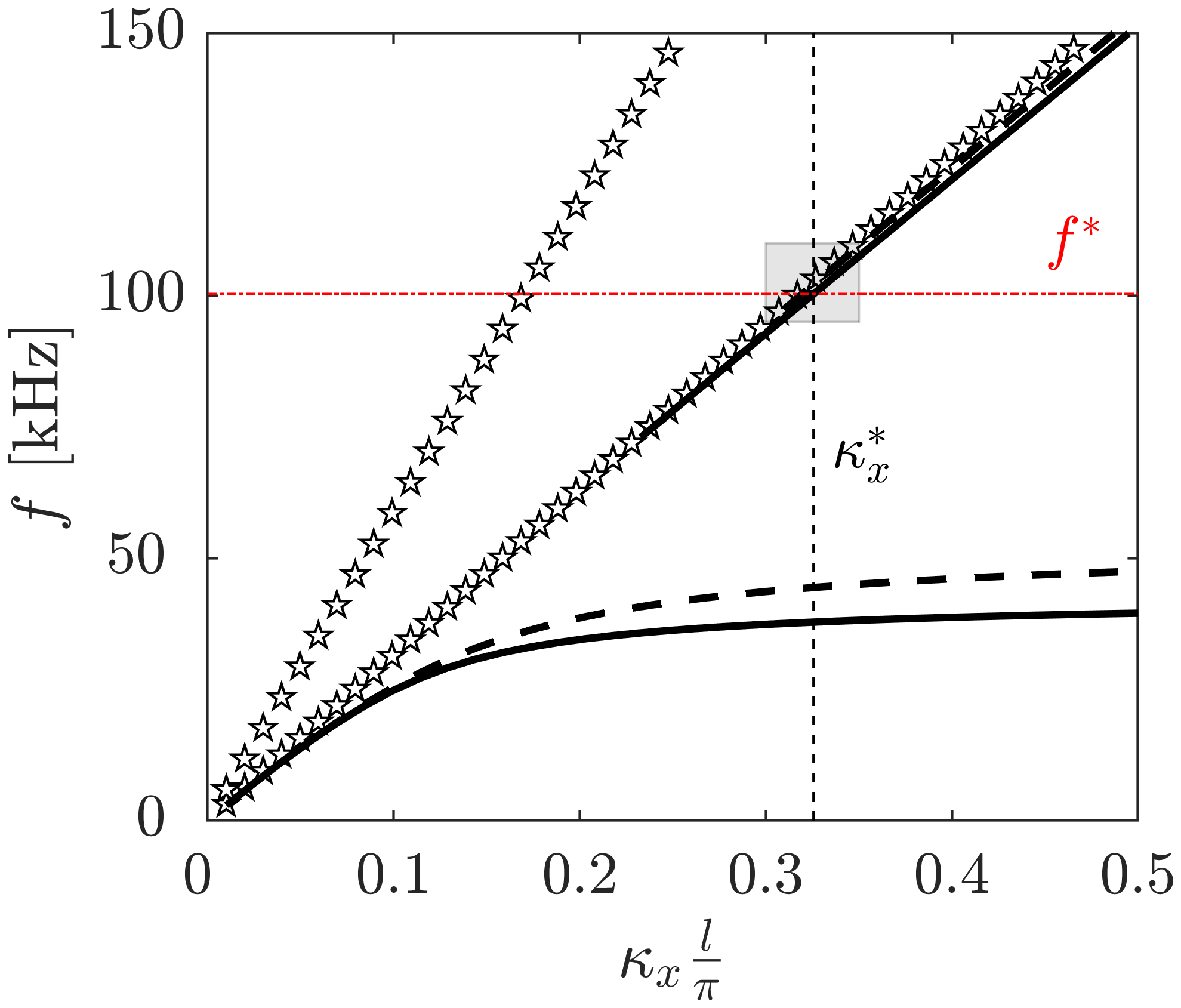}}\quad\hspace{-0.4cm}
        \subfigure[\label{fig: adiabatic case dispersion - zoom}]{\includegraphics[width=.33\textwidth]{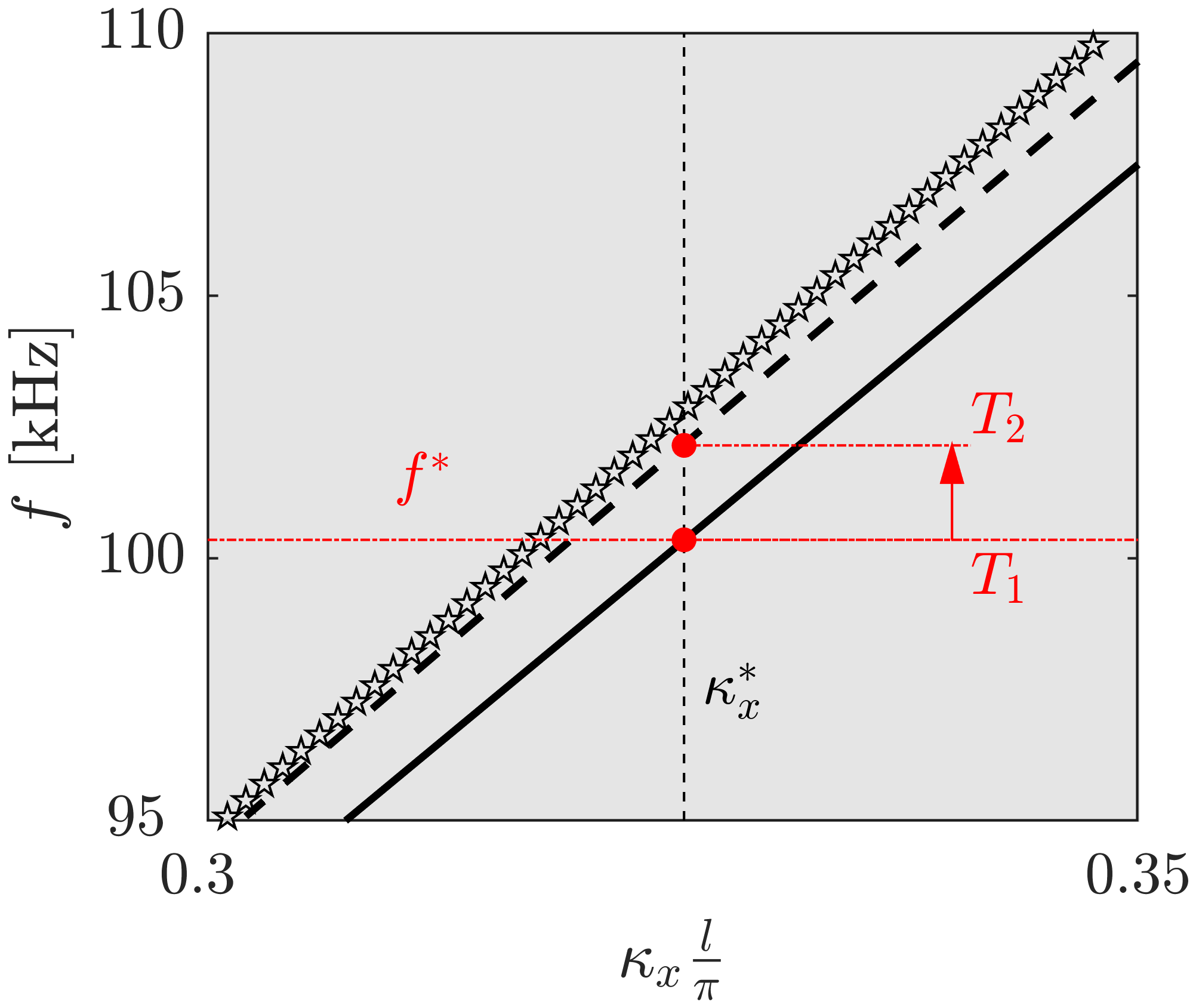}}\quad\hspace{-0.2cm}
        \subfigure[\label{fig: adiabatic limit}]{\includegraphics[width=.32\textwidth]{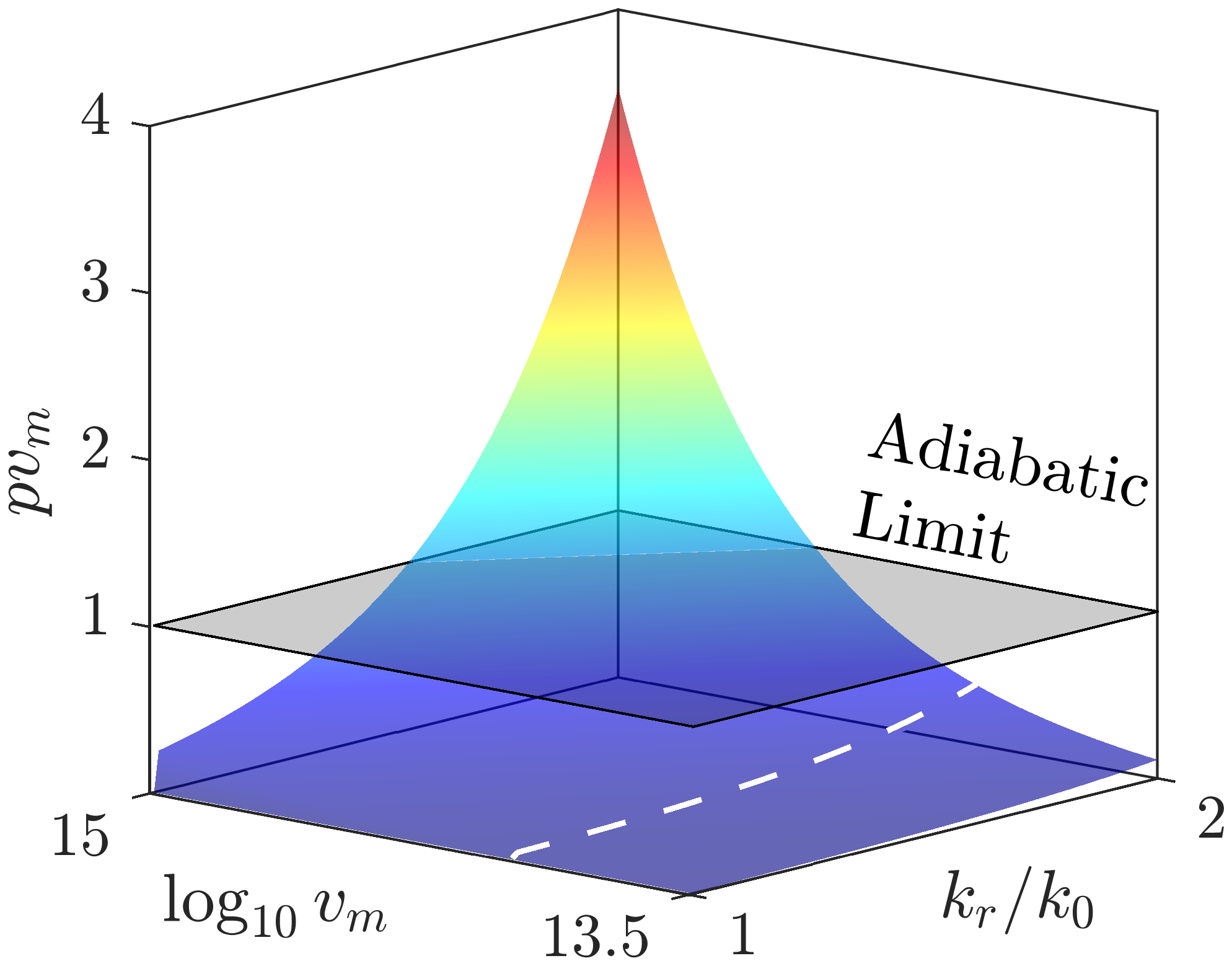}}
        \vfill
        \subfigure[\label{fig: comsol adiabatic dispersion before modulation}]{\includegraphics[width=.33\textwidth]{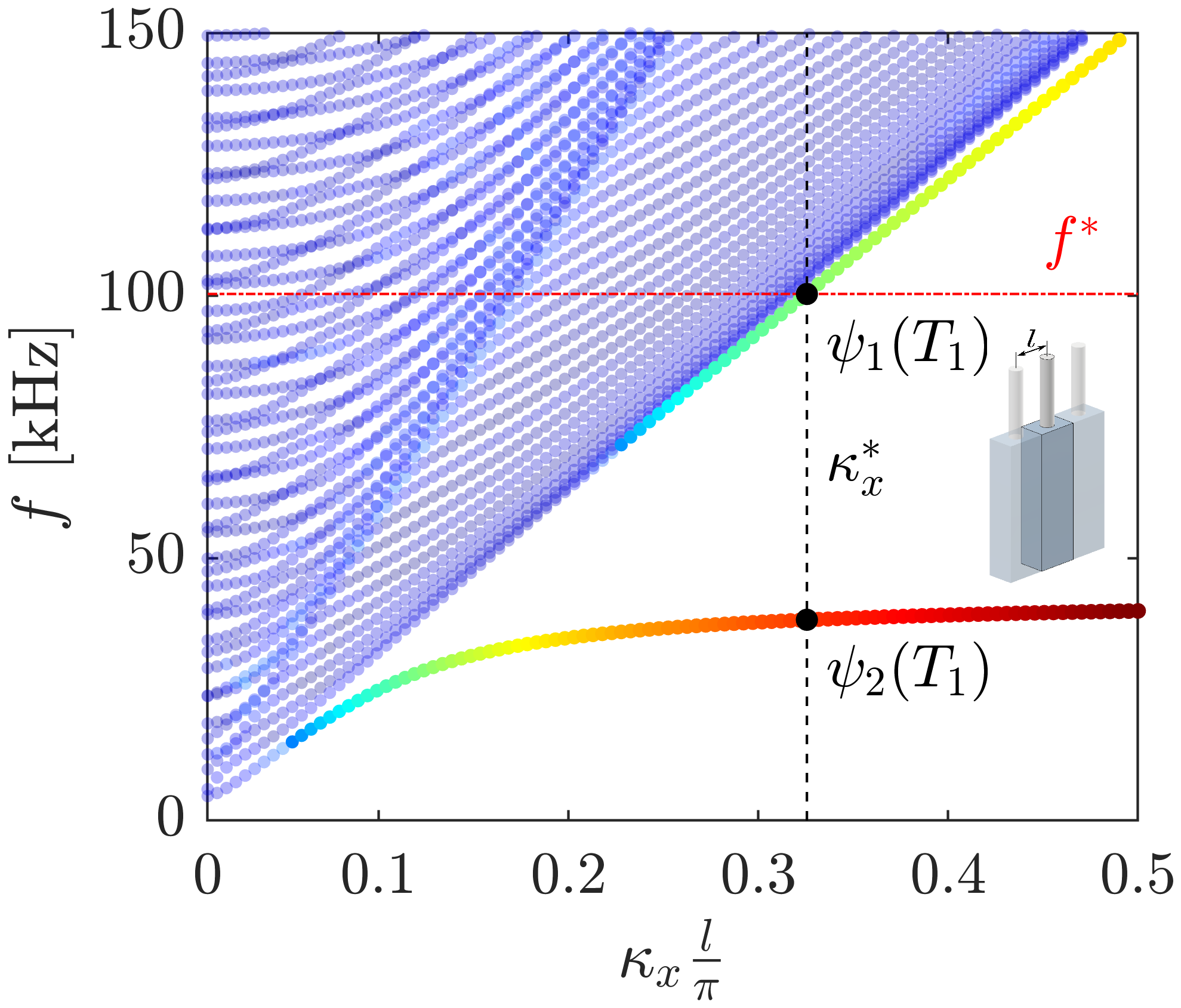}}\quad\hspace{-0.4cm}
        \subfigure[\label{fig: comsol adiabatic dispersion after modulation}]{\includegraphics[width=.33\textwidth]{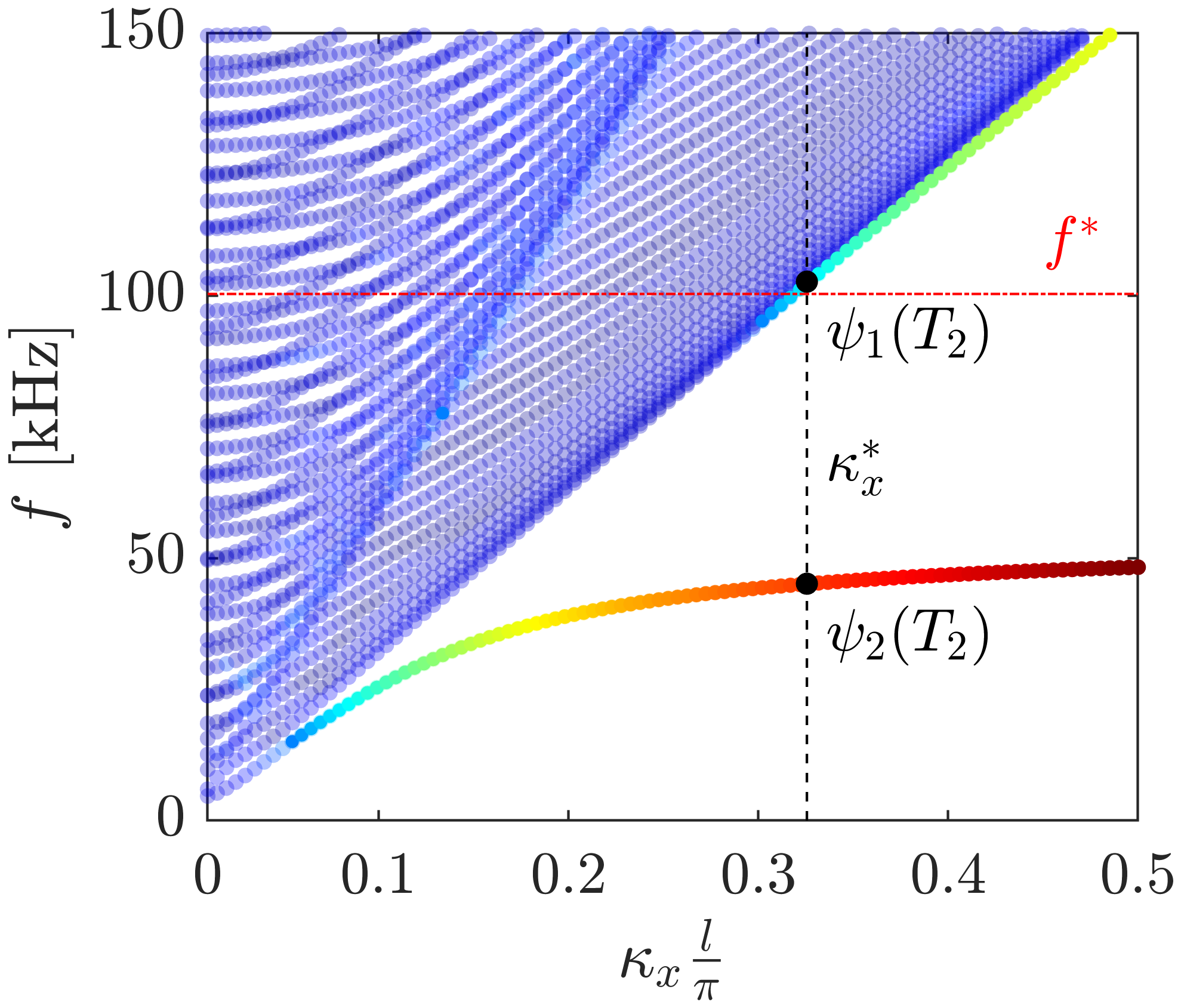}}\quad\hspace{-0.4cm}
        \subfigure[\label{fig: omega varying adiabatic case}]{\includegraphics[width=.335\textwidth]{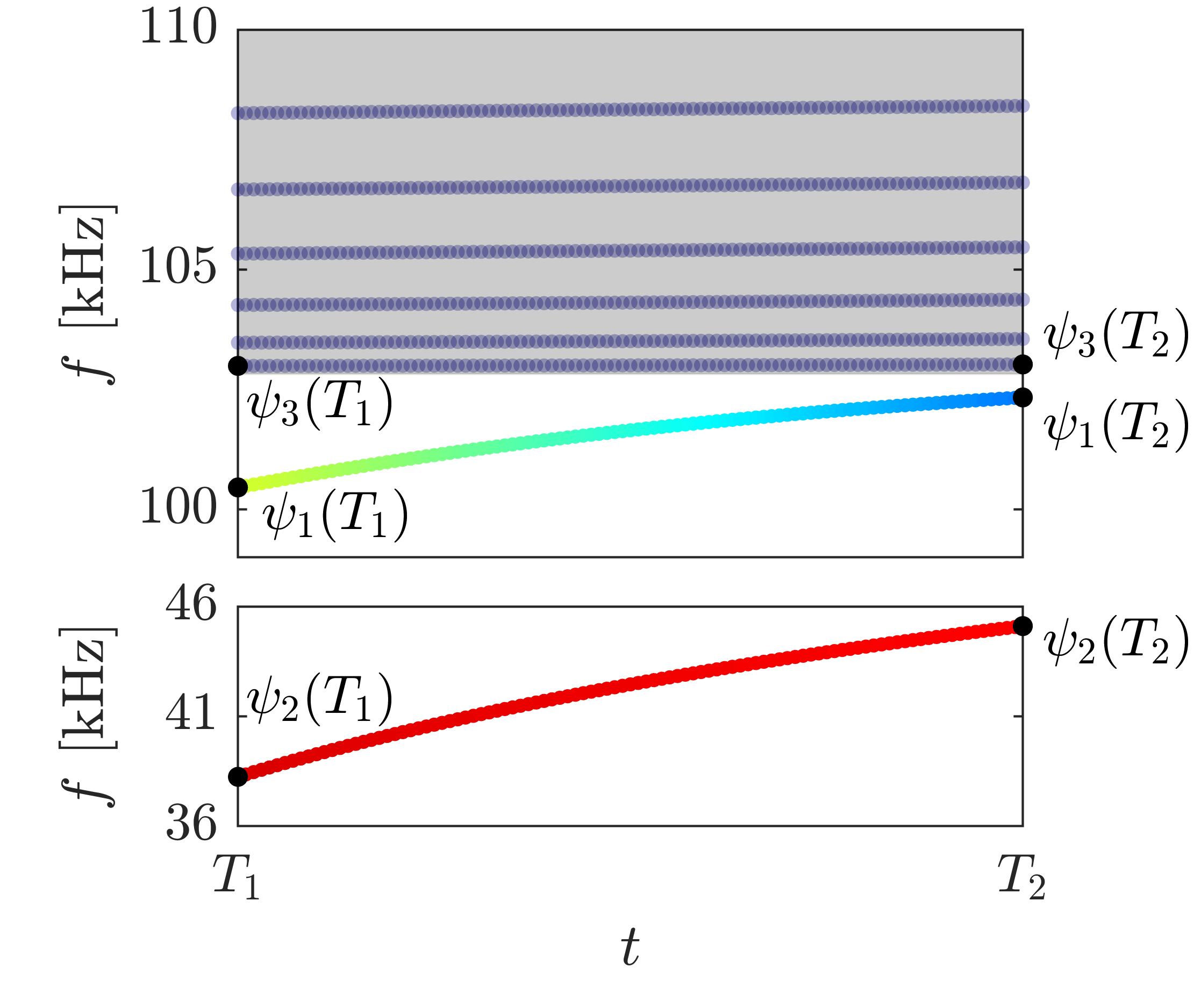}}\quad
         \caption{\subref{fig: adiabatic case dispersion} Analytical dispersion relation obtained upon varying the resonator stiffness from $k_{r1}$ (continuous black lines) to $k_{r2}$ (dashed black lines). Stars are representative of pressure and shear waves. The vertical dashed line denotes the incident wavenumber $\kappa_x^*$ employed (and preserved) during numerical simulations, while the red dash-dotted line represents the frequency of the impinging wave. \subref{fig: adiabatic case dispersion - zoom} Zoomed view of Fig. \subref{fig: adiabatic case dispersion}. \subref{fig: adiabatic limit} Degree of adiabaticity as a function of the modulation speeds, the modulation stiffness, and for the impinging wavenumber $\kappa_x^*$. Velocity and stiffness values generating a curve below the grey horizontal plane can be considered adiabatic. The white dashed line represents a set of parameters used in this paper for simulation purposes. \subref{fig: comsol adiabatic dispersion before modulation}-\subref{fig: comsol adiabatic dispersion after modulation} Numerical dispersion relations computed in a COMSOL Multiphysics environment \subref{fig: comsol adiabatic dispersion before modulation} before and \subref{fig: comsol adiabatic dispersion after modulation} after the modulation takes place. The colors are proportional to the inverse participation ratio (IPR), which denotes localized (red) and bulk modes (blue). The inset displays the unit cell with implied periodic conditions employed in the simulations. \subref{fig: omega varying adiabatic case} Evolution of incident state $\mathbf{\psi}_1\left(\kappa_x^*\right)$ and all neighboring states $\mathbf{\psi}_2\left(\kappa_x^*\right),\;\mathbf{\psi}_3\left(\kappa_x^*\right),\dots$ as time elapses. The grey-shaded region delimits the region populated by bulk modes.}
        \label{fig: dispersion, frequency variation and adiabatic limit}    
    \end{minipage}
\end{figure}

\begin{figure}[t!]
    \centering
    \includegraphics[width=1\textwidth]{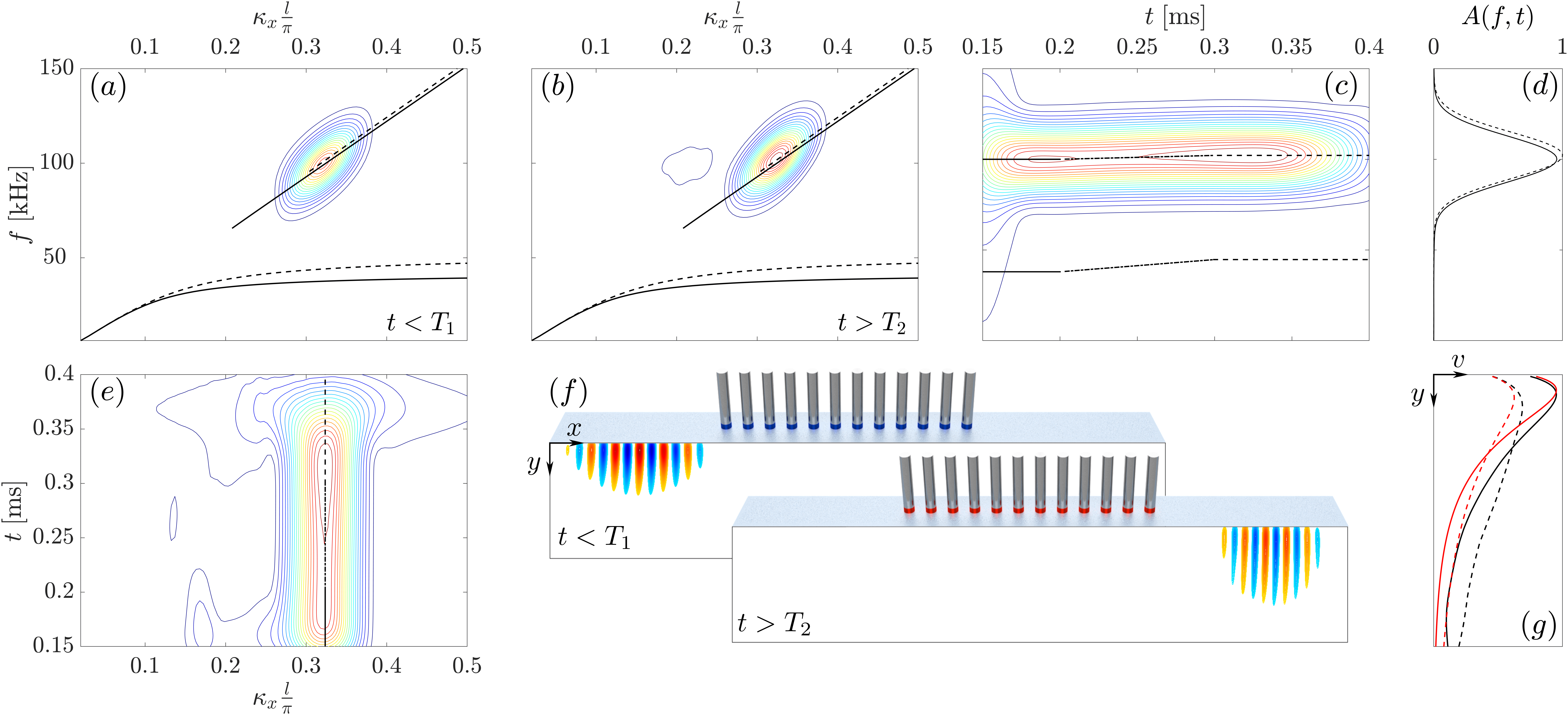}
    \caption{Simulation results for a surface wave impinging an array of time-modulated resonators, where the modulation parameters are compliant with the adiabatic theorem. (a - b) Comparison between the expected and numerical dispersions obtained (a) before ($t < T_1$) and (b) after the modulation ends ($t> T_1$). (c) Frequency spectrogram obtained at constant wavenumber $\kappa_x^*$. The black lines show the expected central frequencies before (continuous line), during (dot-dashed line), and after (dashed line) the modulation. We observe that all the energy remains in correspondence of the excited state, which is the blueprint of adiabaticity. (d) Comparison between the frequency content before and after the modulation. (e) Wavenumber spectrogram for $\kappa_x$. (f) Snapshots of the wavefield before and after time modulation. The wave packet exhibits delocalization. (g) Numerical (black) and analytical (red) surface wave profiles before (continuous line) and after (dashed line) the modulation takes place.}
    \label{fig: adiabatic modulation dynamic results}
\end{figure}
We start the discussion with the schematic illustrated in Fig. \ref{fig: concept}, which consists of a passive solid equipped with an active metasurface specifically engineered to interact with surface acoustic waves (SAW) propagating in the $x-y$ plane. The interaction is achieved through a periodic arrangement of time-modulated mechanical resonators, where the time dependence is herein induced in the form of stiffness-modulation. Note that a mass-modulation could be also used. 
To elucidate the role of time modulation, we illustrate how an impinging wave with frequency $\omega_1$ is transformed over time. The output, dictated by a temporal version of Snell's law, is two-fold and depends on the modulation parameters. First and foremost, this transformation preserves the wavenumber content along the propagation direction while, due to time modulation, the frequency changes to $\omega_2$ \cite{santini2023elastic}. The modulation is hence reminiscent of a knob, which is screwed to vary the frequency content of the output. When the knob is further rotated to produce a greater output frequency $\omega_3$, the process is accompanied by mode conversion from a wave sitting on the surface to a shear wave propagating toward the interior with an inclination $\gamma$, which is tailorable to a certain extent.

The dynamics of such a system is herein studied under the assumption of a quasi-static evolution of the wave modes and is accomplished starting from a dispersion analysis. Details are reported in the supplementary material (SM) \cite{SM}. The computation is carried out following the approach reported in Ref. \cite{palermo2020surface} and the results are illustrated in Fig. \ref{fig: adiabatic case dispersion}. Two stiffness levels $k_{r1}$ and $k_{r2}$ are herein considered, which correspond to the initial and final stiffness values represented with solid and dashed lines. Stars are representative of pressure and shear waves with characteristic speeds $c_L$ and $c_S$, respectively.

We now consider a configuration where the effective stiffness is initially $k_{r1}=k_0$ and smoothly evolves to $k_{r2}=1.57\;k_0$ following a linear modulation $k(t) = k_{r1}+v_m(t-T_1)$, where $v_m$ is the velocity of the modulation.

The energy, initially injected in the neighborhood of the dispersion branch at $f^* = 100$ kHz, must follow the wavenumber-preserving transformation illustrated with a vertical dashed line in Fig. \ref{fig: adiabatic case dispersion} which, to ease visualization, is also reported in the zoomed view (Fig. \ref{fig: adiabatic case dispersion - zoom}). In both diagrams, $\kappa_x^*$ is the impinging wavenumber corresponding to $f^*$ and $\kappa_x^*l/\pi$ is its dimensionless form. $l$ is the lattice size.
Provided the modulation is slow enough, the spectral content of the wave is expected to adiabatically follow the Rayleigh wave dispersion curve and, hence, induce a frequency transformation from the solid line to the dashed line. It is initially assumed that the stiffness modulation is sufficient only to induce a small shift in frequency and, hence, without triggering any interaction between Rayleigh and shear modes, which would be present in case the Rayleigh dispersion approaches the shear dispersion \cite{colombi2016seismic}.

We remark that the analysis herein discussed is relative to a wave propagating along $x$ with evanescent wavenumber $\kappa_y$, which determines the degree of penetration of the wave rather than an inclined propagation toward the bulk. We also remark that the shear and pressure waves displayed in Figs. \ref{fig: adiabatic case dispersion} and \ref{fig: adiabatic case dispersion - zoom} are characterized by $\kappa_y=0$, since waves with $\kappa_y\neq0$ would populate another plane of the dispersion cone $\omega\left(\kappa_x,\kappa_y\right)$. This concept will be further discussed in the following section and is key for mode conversion. 

To shed light on the adiabaticity of this transformation, the dispersion analysis of a finite strip of half-plane with implied $x$-periodic conditions is developed in a COMSOL Multiphysics environment. The dimension along $y$ is set sufficiently large to emulate an infinite medium. To ease readability, details on the dispersion analysis, methods, and system parameters are reported in the supplementary material (SM) \cite{SM}. 
Similar to previous studies on elastic and acoustic lattices \cite{santini2023elastic,riva2023adiabatic2}, the framework herein discussed allows us to discriminate between adiabatic and nonadiabatic modulations which, depending on the modulation velocity, may drive the energy transfer between the incident Rayleigh wave and all other bulk and surface wave modes populating the solid for the impinging wavenumber. 

We now focus on the numerical dispersion relation shown in Fig. \ref{fig: comsol adiabatic dispersion before modulation} and Fig. \ref{fig: comsol adiabatic dispersion after modulation} obtained with $k_{r1} = k_0$ and $k_{r2} = 1.57\;k_0$, respectively. The diagrams are color-coded: colors are proportional to the inverse participation ratio (IPR), which quantitatively describes mode localization and is defined as:
\begin{equation}
    IPR = \frac{\int_S\;\left<\mathbf{\psi}_n|\mathbf{\psi}_n\right>^2 dS}{\left(\int_S\left<\mathbf{\psi}_n|\mathbf{\psi}_n\right> dS\right)^2}
    \label{eq: ipr}
\end{equation} 
where $S$ is the unit cell domain. $\left|\mathbf{\psi}_n\right>$ is the eigenvector relative to the eigenvalue $\omega_n$ evaluated through the reduced eigenvalue problem $j\omega\left|\mathbf{\psi}_n\right>=H\left(\kappa_x,t\right)\left|\mathbf{\psi}_n\right>$, where the Hamiltonian matrix $H\left(\kappa_x^*,t\right)$ quasi-statically varies over time.
Red dots are representative of localized modes, e.g. a surface wave, while blue dots denote a bulk wave. Note that the hybridized Rayleigh wave is highly localized near resonance, while the upper branch becomes less localized as the Rayleigh dispersion approaches the bulk one. This is key for the following discussion.

We now discuss how the dispersion of the strip for the impinging wavenumber $\kappa_x^*$ varies during time modulation. Results are reported in Fig. \ref{fig: omega varying adiabatic case}. The evolution of all wave modes is shown, including those populating the bulk and herein delimited within the grey area. We observe that these modes do not change their frequency and localization properties. In contrast, Rayleigh waves experience a frequency transformation accompanied by a delocalization mechanism whereby the degree of penetration changes over time. 
In this section, we stop the knob before the surface mode enters the grey region. This limitation will be later relaxed to show how a larger stiffness modulation allows driving a conversion mechanism between the surface wave and the bulk modes.

We remark that Fig. \ref{fig: omega varying adiabatic case} is representative of a quasi-static (adiabatic) transformation of an incident state, i.e. in case the modulation from $k_{r1}$ to $k_{r2}$ is sufficiently slow for the energy not to be scattered towards other wave modes. We now assess the adiabaticity of this transformation under the assumption of an arbitrary modulation speed and by following the procedure in Ref. \cite{santini2023elastic,xia2021experimental,riva2023adiabatic}, which requires the elastodynamic equation to be written in terms of first-order differential form akin to Schrödinger's equation for an impinging wavenumber $\kappa_x^*$:
\begin{equation}
\left|\mathbf{\Psi}_n\right>_{\;,t}=H\left(\kappa_x^*,t\right)\left|\mathbf{\Psi}_n\right>
\end{equation}
The energy, initially injected as a Rayleigh wave, is represented as $\left|\mathbf{\Psi}_1\right>$ at $t=T_1$ in Fig. \ref{fig: omega varying adiabatic case}. The transformation, which takes place over time, can be considered adiabatic if there is negligible energy leakage toward the neighboring states $\left|\mathbf{\Psi}_2\right>$, $\omega_2$ or, alternatively, $\left|\mathbf{\Psi}_3\right>$, $\omega_3$. We found $\left|\mathbf{\Psi}_2\right>$, $\omega_2$ more critical from the adiabaticity viewpoint, which is justified by the smaller frequency separation $\omega_2-\omega_1$. For adiabaticity, the following inequality must be satisfied:
\begin{equation}
    p\;v_m << 1\hspace{1cm}where:\hspace{0.5cm}p=\left|\frac{\Big<\mathbf{\psi}_2^L\left|\displaystyle\frac{\partial H}{\partial k_r}\right|\mathbf{\psi}_1^R\Big>}{\left(\omega_2-\omega_1\right)^2}\right|
    \label{eq: adiabatic limit}
\end{equation}
The term on the left-hand side is computed for the impinging wavenumber $\kappa_x^*$ and mapped in Figure \ref{fig: adiabatic limit} as a function of time and upon varying the modulation speed $v_m$. For sufficiently small modulation speeds, the transformation can be considered adiabatic, i.e. when the surface is below the unitary threshold highlighted by the grey plane. In the diagram, the white-dashed curve is relative to the adiabatic modulation employed for simulation purposes and hereafter discussed in detail.

\begin{figure}[t!]
    \centering
    \includegraphics[width=\textwidth]{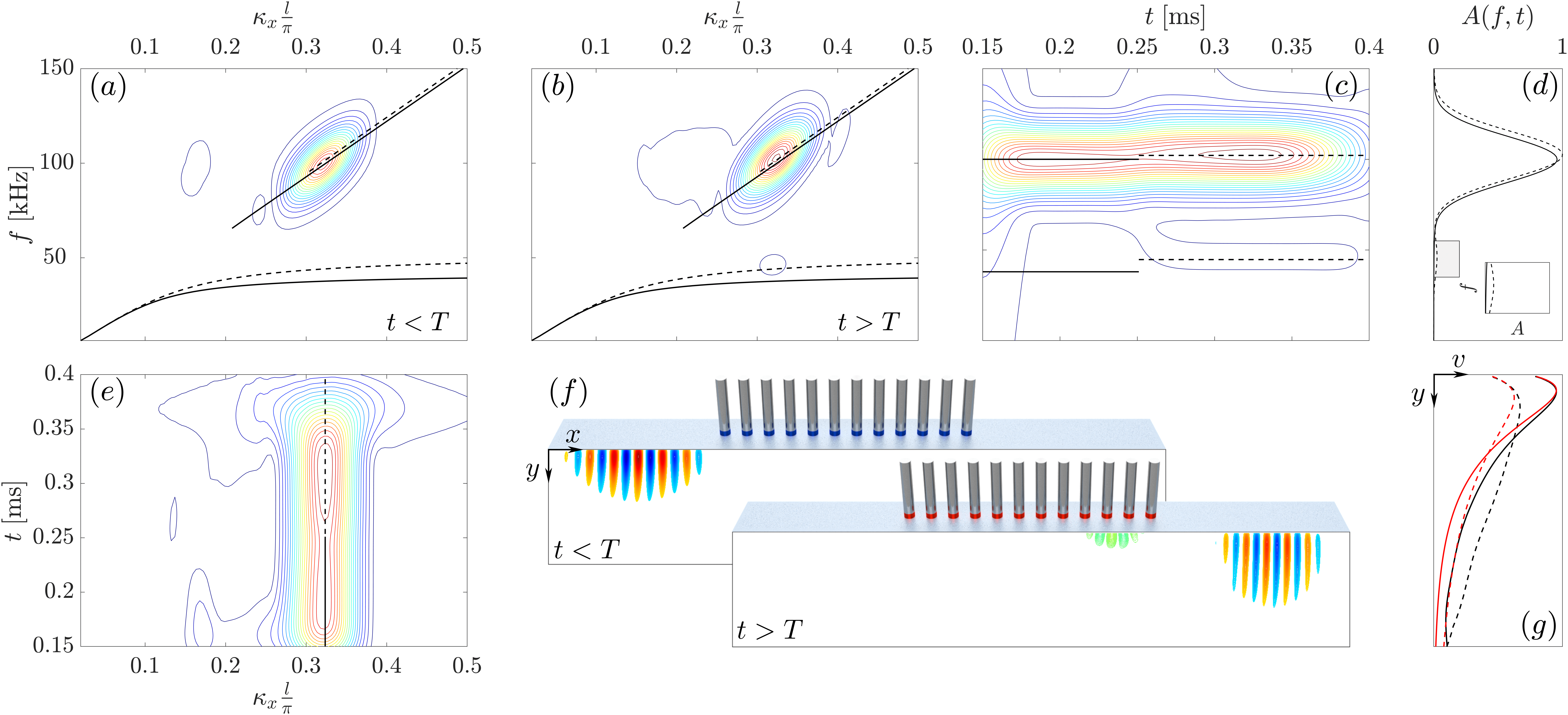}
    \caption{Simulation results for a surface wave impinging on an array of resonators modulated with a non-adiabatic (very fast) transformation in time. (a)-(b) Numerical dispersion before and after the step modulation centered at $\tau = T$. (c) Frequency spectrogram at constant wavenumber $\kappa_x^*$. Black lines show the expected central frequency before (continuous line) and after (dashed line) the modulation. The energy, initially injected in correspondence of the excited state, leaks toward all the states that populate the dispersion relation after modulation. (d) Comparison between the frequency content before and after the modulation. In the inset, a zoomed view of the shaded area is used to highlight leakage towards the lower branch of the dispersion. (e) Wavenumber spectrogram for $\kappa_x$. (f) Wavefield before and after time modulation. The wave packet exhibits delocalization. An additional mode, caused by the non-adiabaticity of the process, is present on the surface. (g): Numerical (black) and analytical (red) wave profiles before (continuous line) and after (dashed line) the modulation.}
    \label{fig: step modulation with delocalization}
\end{figure}

The theoretical discussion is now verified through numerical simulations. To delineate the transition between adiabatic and non-adiabatic modulations, two scenarios with the same starting and ending stiffness values, but different modulation speeds, are reported. Fig. \ref{fig: adiabatic modulation dynamic results} illustrates results compliant with the adiabatic limit, whereby the velocity of modulation is low. In contrast, Fig. \ref{fig: step modulation with delocalization}, illustrates the results with a step-like modulation, which can be considered nonadiabatic. 

Two representative snapshots before and after time modulation are reported in Fig. \ref{fig: adiabatic modulation dynamic results}(f). As expected, since the modulation is adiabatic, the wave packet experiences a transformation without energy leak toward other modes populating the surface and the bulk material. As a result, a single wave packet is obtained as output, which is however characterized by a greater degree of penetration in the bulk of the material, as anticipated through the numerically computed IPR in Figs. \ref{fig: omega varying adiabatic case}, \ref{fig: comsol adiabatic dispersion before modulation} and \ref{fig: comsol adiabatic dispersion after modulation}.
The analytical (red) and numerical (black) envelopes before (continuous line) and after (dashed line) the modulation are also reported in Fig. \ref{fig: adiabatic modulation dynamic results}(g) for completeness.

The time history is further processed to shed more light on the results, as shown in Fig. \ref{fig: adiabatic modulation dynamic results}(a)-(b). The figure compares the analytical and numerical (reconstructed from simulation) dispersions of the medium in reciprocal space before ($t<T_1$) and after ($t>T_2$) the modulation takes place. 
The numerical dispersion is computed by taking the 3D Fourier transform of wave motion $\bm{u}\left(x,y,t\right)$ measured within the solid, which yields the complex-valued coefficients $\bm{\hat{u}}\left(\kappa_x,\kappa_y,\omega\right)$. The wavenumber dependence $\kappa_y$ is eliminated by taking the RMS value along $\kappa_y$, leading to the dispersion $\bm{\hat{u}}\left(\kappa_x,\omega\right)$ displayed as a contour plot. 
In the first diagram, the energy content follows the solid line, which is representative of the dispersion with $k_{r1}$. In the second diagram, the energy content matches the dashed line, which is relative to the final modulation value $k_{r2}$. Note that the amplitude is zero elsewhere and, hence, we can conclude that a scattering-free frequency conversion is achieved. 

The frequency shift is more visible in the spectrogram reported in Fig. \ref{fig: adiabatic modulation dynamic results}(c), which is obtained by windowing the time history with a moving Gaussian function $e^{-\left(t-t_0\right)^2/2c^2}$. Here, the time history is Fourier-transformed to $\hat{u}\left(\kappa_x,\kappa_y,\omega,t_0\right)$, and reduced to $\hat{u}\left(\omega,t_0\right)$ by taking the RMS value along both wavenumber dimensions. Also, the spectrum $\hat{u}\left(\omega,t<T1\right)$ and $\hat{u}\left(\omega,t>T2\right)$, evaluated before and after time modulation, is reported alongside the spectrogram (see Fig. \ref{fig: adiabatic modulation dynamic results}(d)), which further shows the aforementioned frequency shift.
We observe that, due to the adiabatic nature of the modulation, the incident wave evolves sufficiently slow for the transformation to avoid scattering toward the other available wave modes, and, as such, the energy remains successfully located in the upper state. Finally, the spectrogram reported in Fig. \ref{fig: adiabatic modulation dynamic results}-V demonstrates the conservation of the wavenumber content, which is preserved during time modulation. 

The outcome of a non-adiabatic transformation, obtained through a very fast modulation of the stiffness, is displayed in Fig. \ref{fig: step modulation with delocalization}. Interestingly, the envelopes before and after time modulation (Fig. \ref{fig: step modulation with delocalization}(g)) are very similar to the adiabatic case, but there is an additional wave packet that nucleates due to the temporal modulation, which is visible in Fig. \ref{fig: step modulation with delocalization}(f) for $t>T$. 
To shed light on the nature of this additional wave packet and the underlying dynamics, the spectral content of the simulation is reported in Figs.\ref{fig: step modulation with delocalization}(a)-(c). 
In contrast to the adiabatic case, we observe a bit of energy leak toward the resonant branch of the dispersion. This is particularly visible in the frequency spectrogram in Fig.\ref{fig: step modulation with delocalization}(c), where the frequency conversion is accompanied by an energy jump, also present in Fig.\ref{fig: step modulation with delocalization}(d). This energy jump is justified by the additional wave packet formed across the time interface. 

\section{Mode conversion induced by slow and fast time modulation}\label{sec3}
\begin{figure}[t!]
    \centering
    \begin{minipage}{1\textwidth}
        \centering
        \hspace{-1cm}\subfigure[\label{fig: mode conversion case dispersion}]{\includegraphics[width=.33\textwidth]{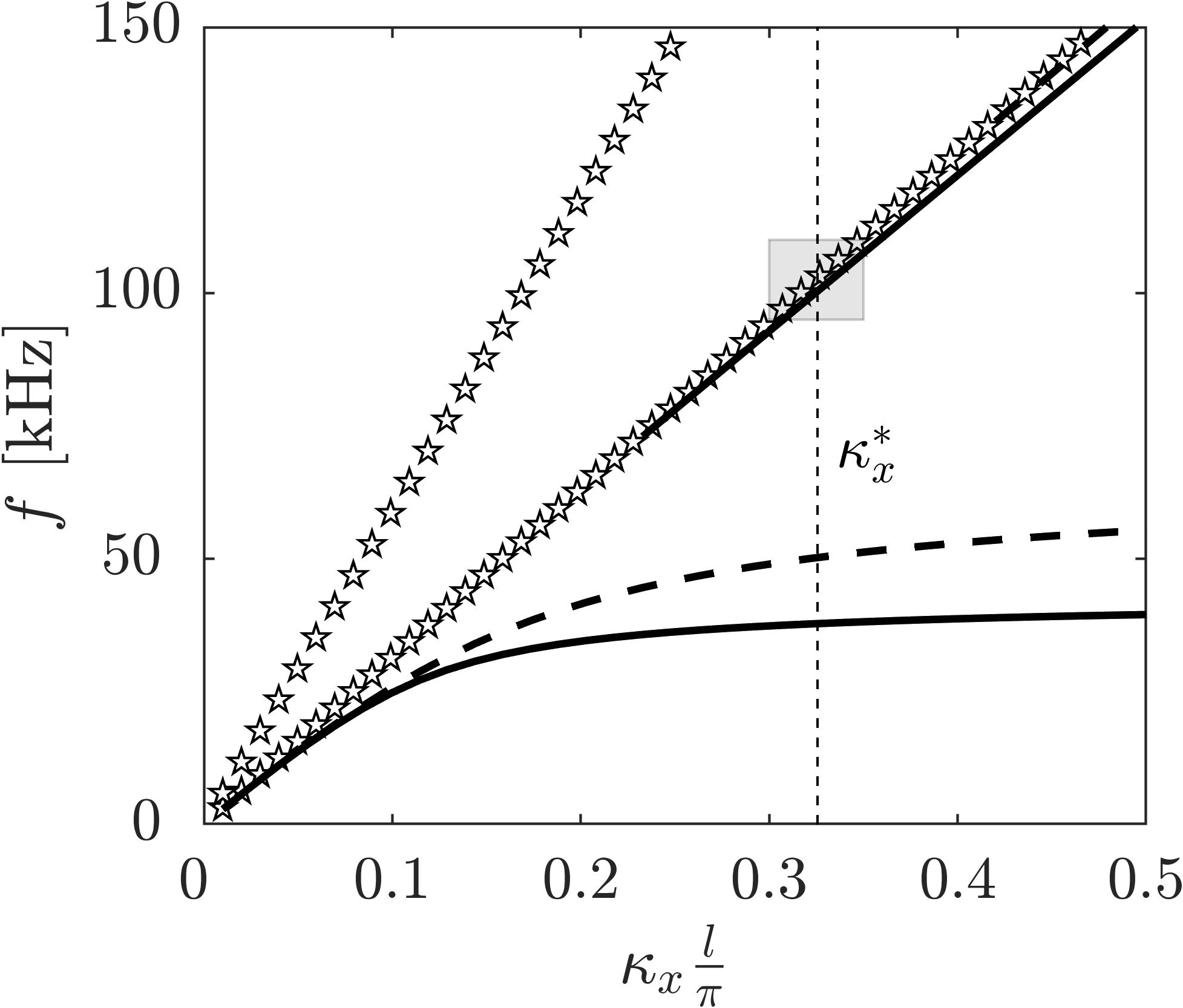}}
        \subfigure[\label{fig: mode conversion case dispersion - zoom}]{\includegraphics[width=.33\textwidth]{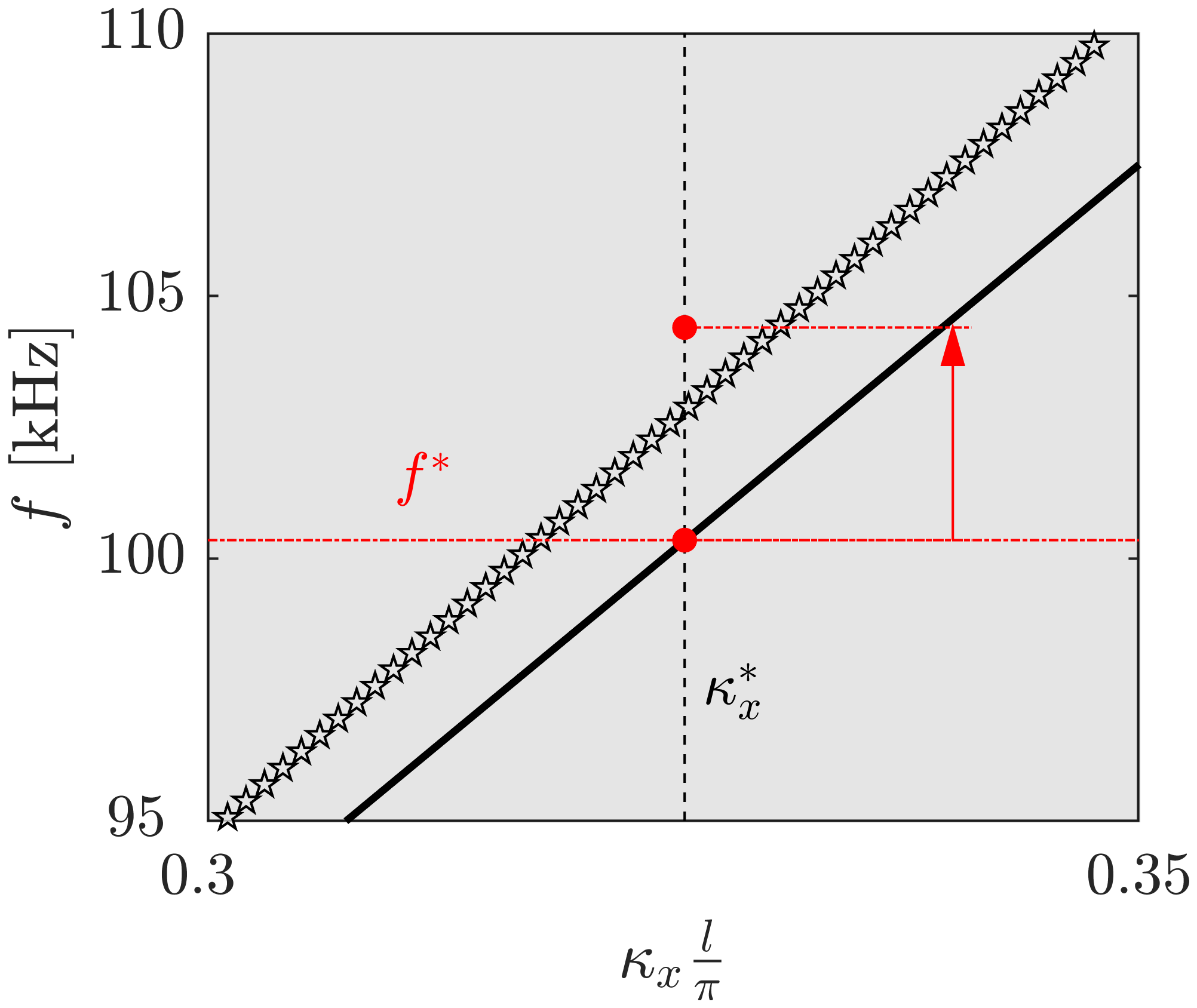}}
        \caption{\subref{fig: mode conversion case dispersion} Dispersion relation upon varying the resonator stiffness from $k_{r1}$ (continuous black lines) to $k_{r3}$ (dashed black lines). White stars represent the pressure and shear modes. The vertical dashed line represents the incident wavenumber $\kappa_x^*$, which is preserved during the modulation. \subref{fig: mode conversion case dispersion - zoom} Zoomed view of Fig. \subref{fig: mode conversion case dispersion} limited to the grey highlighted area. The arrow illustrates that the wavenumber-preserving transformation is tailored to frequency-convert the impinging energy above the shear cone.}
        \label{fig: dispersion, frequency variation and mode conversion}
    \end{minipage}
\end{figure}
\begin{figure}[h!]
    \centering
    \includegraphics[width=\textwidth]{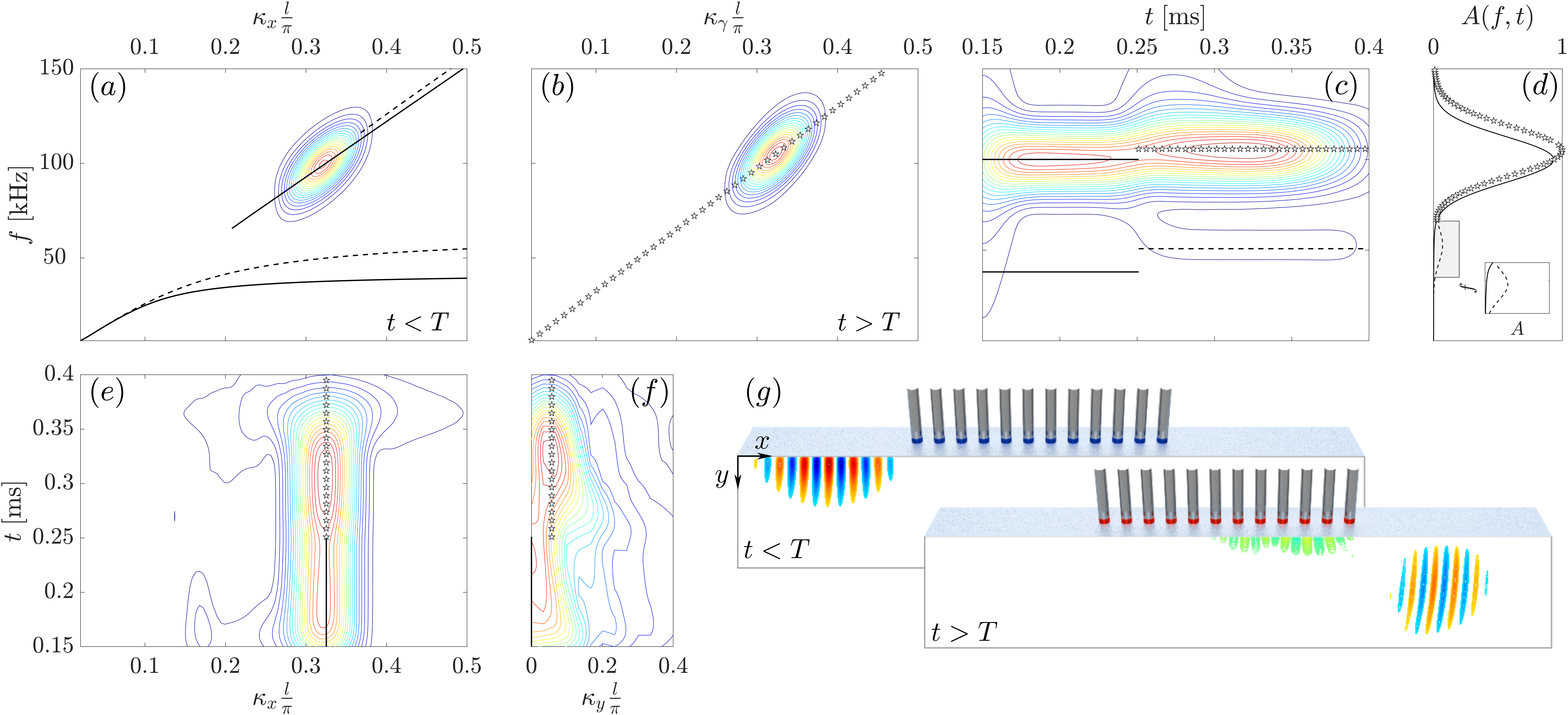}
    \caption{Simulation results for a surface wave impinging on an array of resonators modulated with a non-adiabatic (very fast) transformation in time. (a)-(b) Numerical dispersion relation before and after the step modulation at time $t = T$. (c) Frequency spectrogram at constant wavenumber $\kappa_x^*$. Black lines show the expected central frequencies of the surface waves before (continuous line) and after (dashed line) the modulation. White stars represent the output frequency of the shear mode for the impinging wavenumber $\kappa_x^*$ achieved after the modulation. (d) Comparison between the frequency content before and after the modulation. The inset highlights leakage towards the lower branch of the dispersion. (e)-(f) Wavenumber spectrogram for (e) $\kappa_x$ and (f) $\kappa_y$. (g) Wavefield before and after the modulation. The wave packet exhibits surface-to-bulk mode conversion and additional modes nucleate on the surface due to the non-adiabatic nature of the process.}
    \label{fig: mode conversion dynamic results pt1}
\end{figure}

\begin{figure}[h!]
    \centering
    \hspace{-0.1cm}\includegraphics[width=1\textwidth]{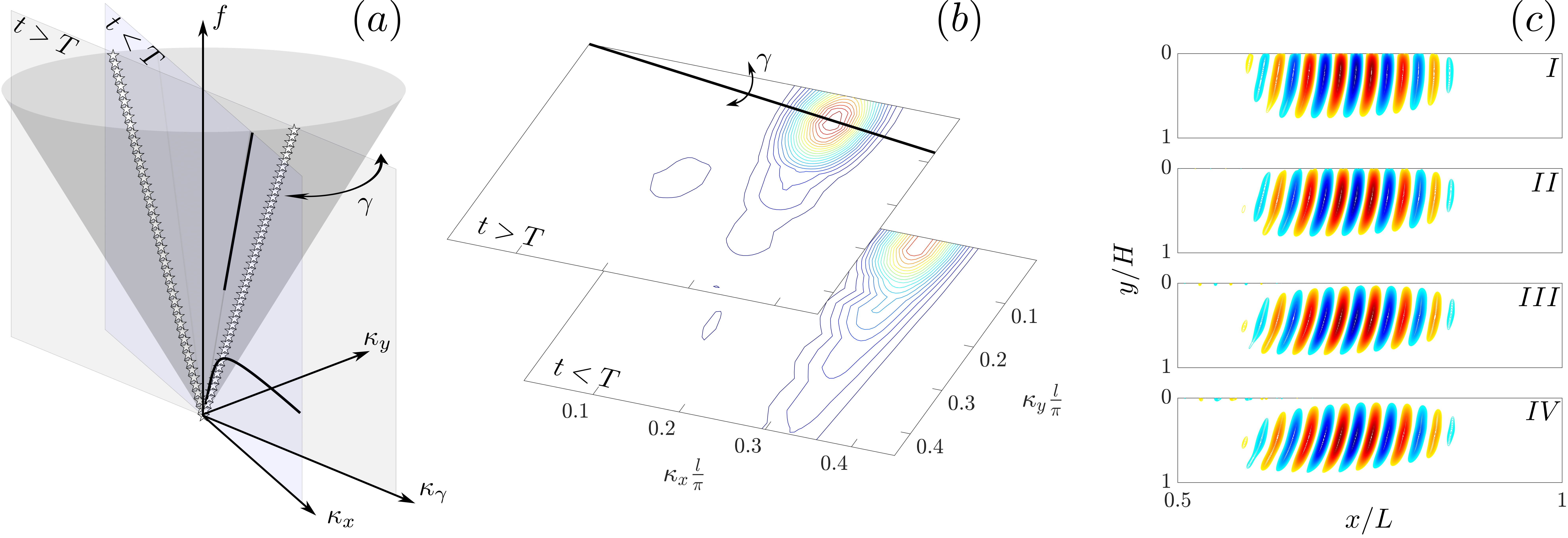}
    \caption{(a) Orientation of the planes hosting the dispersion in case of mode conversion. The grey cone represents shear modes and the blue plane hosts the Rayleigh dispersion. $\kappa_\gamma-\omega$ is the plane after the modulation takes place. Waves populating the $\kappa_\gamma-\omega$ plane point towards the bulk of the material with an inclination $\gamma$. (b): Numerically reconstructed wavenumber content before and after the modulation. There is a visible change of $\kappa_y$ that occurs over time and is responsible for a non-null $\gamma$. (c) Snapshots of the wavefield for $t>T$ evaluated for different stiffness modulation values $k_r$, (I) $k_r = 1.95\;k_0$, (II) $k_r = 2.40\;k_0$, (III) $k_r = 2.91\;k_0$, and (IV) $k_r = 3.25\;k_0$. The angle at the outlet varies with $k_r$.} 
    \label{fig: mode conversion dynamic results pt2}
\end{figure}

We now further rotate the knob to achieve mode conversion, whereby the stiffness is modulated to shift the upper dispersion branch above the shear cone for the impinging wavenumber $\kappa_x^*$, see Fig. \ref{fig: dispersion, frequency variation and mode conversion}. 
In contrast to prior works on space metagradings, we remark that a gradual variation of the properties would slowly transform the surface mode into a shear wave populating the $\kappa_x-\omega$ plane, hence, without any directionality $\gamma$ toward the interior, which would be dictated by non-evanescent and non-null values of $\kappa_y$. This behavior is justified by the adiabatic nature of the transformation, which implies energy conversion into a bulk mode before a further increase of the frequency content takes place. This assumption is numerically verified through a slow modulation from $k_{r1}=k_0$ to $k_{r3}=2.4\;k_0$, thereby achieving full conversion between a surface wave to a shear wave with $\kappa_y=0$ and, hence, with a null propagation angle $\gamma=0$. Numerical results on this matter are reported in the supplementary material (SM) \cite{SM}.

A nonadiabatic step modulation is hence necessary to suddenly change the frequency content $\omega_1\rightarrow\omega_3$ of the surface wave above the shear cone which, due to momentum conservation along $x$, would determine a complex-valued wavenumber $\kappa_y=\sqrt{{\kappa_x^*}^2-\omega_3^2/c_S^2}$ at the outlet. In contrast to adiabatic modulations with mode conversion, the energy sitting on the surface is altered by the time duration of the drive, which is very quick. Hence, we expect the frequency conversion illustrated by the arrow in Fig. \ref{fig: dispersion, frequency variation and mode conversion}(b) to be accompanied by undesired mode scattering towards the available states that populate the solid. The stiffness levels considered herein are again $k_{r1}=k_0$ and $k_{r3}=2.4\;k_0$, corresponding to the solid and dashed line in the figure, respectively, while stars are representative of shear and pressure waves.

The concept is corroborated via numerical simulations. In contrast to the previous results, three wave modes are visible in Fig. \ref{fig: mode conversion dynamic results pt1}(g) for $t>T$, i.e. after the modulation takes place, which consist of two counter-propagating surface waves and a shear wave. As expected, the impinging wave is converted into a shear wave in response to a frequency transformation $\omega_1\rightarrow\omega_3$ and the non-adiabatic nature of the modulation is revealed by the nucleation of additional wave packets on the surface with minor energy content. 

The frequency shift and the formation of such waves are better illustrated in the frequency domain, see Fig. \ref{fig: mode conversion dynamic results pt1}(a)-(c). First, we note that before and after the modulation, the numerical and theoretical dispersion are in good agreement. However, while Fig. \ref{fig: mode conversion dynamic results pt1}(a) displays results in the $\kappa_x-\omega$ plane, Fig. \ref{fig: mode conversion dynamic results pt1}(b) illustrates that the energy content lies in the $\kappa_\gamma-\omega$ plane, where $\gamma$ is coincident with the expected propagation angle $\gamma=\arctan{\left(\kappa_y/\kappa_x\right)}$. This means that the frequency up-conversion is accompanied by a wavenumber transformation for $\kappa_y$. To ease visualization, the two planes are reported schematically in Fig. \ref{fig: mode conversion dynamic results pt2}(a). In addition, due to the nonadiabatic nature of the modulation, both the frequency spectrogram and spectrum before and after modulation in Fig. \ref{fig: mode conversion dynamic results pt1}(c)-(d) not only report frequency up-conversion, but also energy leakage towards the resonant dispersion branch. This justifies the presence of additional surface modes trapped within the array of resonators. 

 Now, the time history is processed to show the wavenumber content in reciprocal space and relative spectrograms. The resulting wavenumber spectrograms in Fig. \ref{fig: mode conversion dynamic results pt1}(e)-(f) further demonstrate momentum conservation for $\kappa_x^*$, while there is an evolution of wavenumber $\kappa_y$ toward a nonzero value. This \textit{de facto} determines a nonzero angle $\gamma$ visible in the $\kappa_x-\kappa_y$ plane, see Fig. \ref{fig: mode conversion dynamic results pt2}(b), which better highlights the inclination angle achieved in the numerical simulation after the modulation takes place. We remark that further increasing the final stiffness value would modify the wavenumber $\kappa_y=\sqrt{{\kappa_x^*}^2-\omega_3^2/c_S^2}$ and, hence, the angle $\gamma$. This is shown through the temporal snapshots illustrated in Fig. \ref{fig: mode conversion dynamic results pt2}(c), which are obtained with different stiffness modulation values. To ease readability, the spectral analysis is left in the supplementary material (SM) \cite{SM}. 

\begin{figure}
    \centering
    \subfigure[\label{fig: snapshots upper & lower array}]{\includegraphics[width=.5\textwidth]{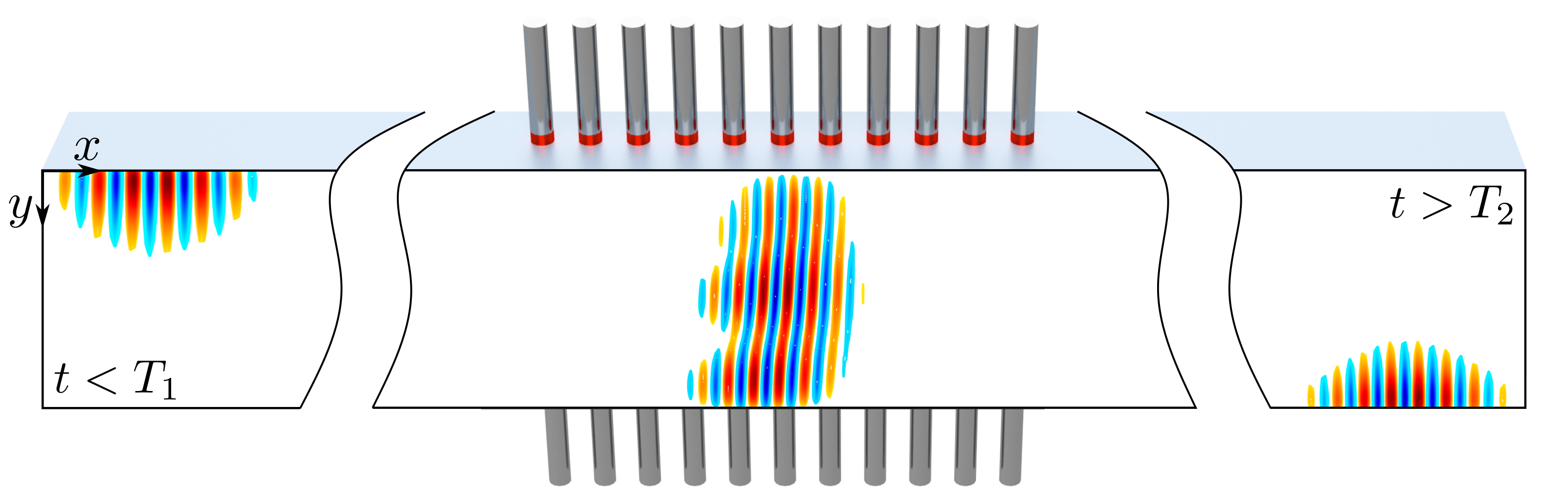}}\quad
    \subfigure[\label{fig: upper & lower array - rms}]{\includegraphics[width=.47\textwidth]{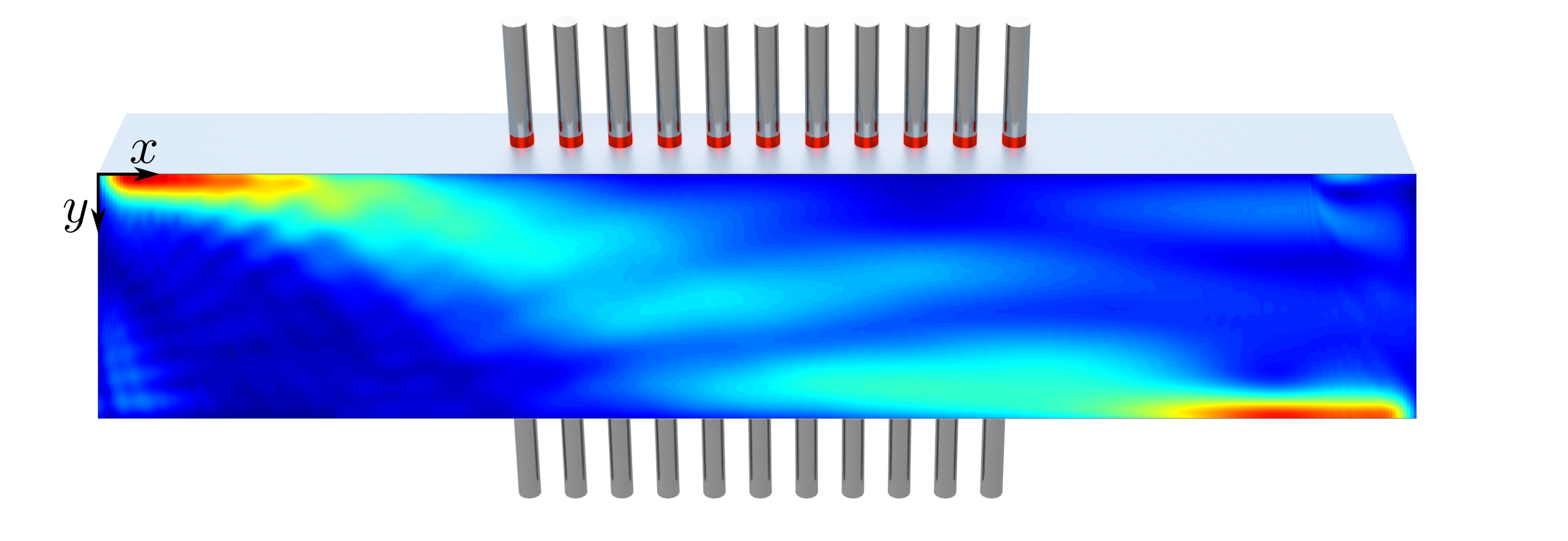}}\vfill
    \hspace{-5pt}\subfigure[\label{fig: upper & lower array - spectrogram}]{\includegraphics[width=.4\textwidth]{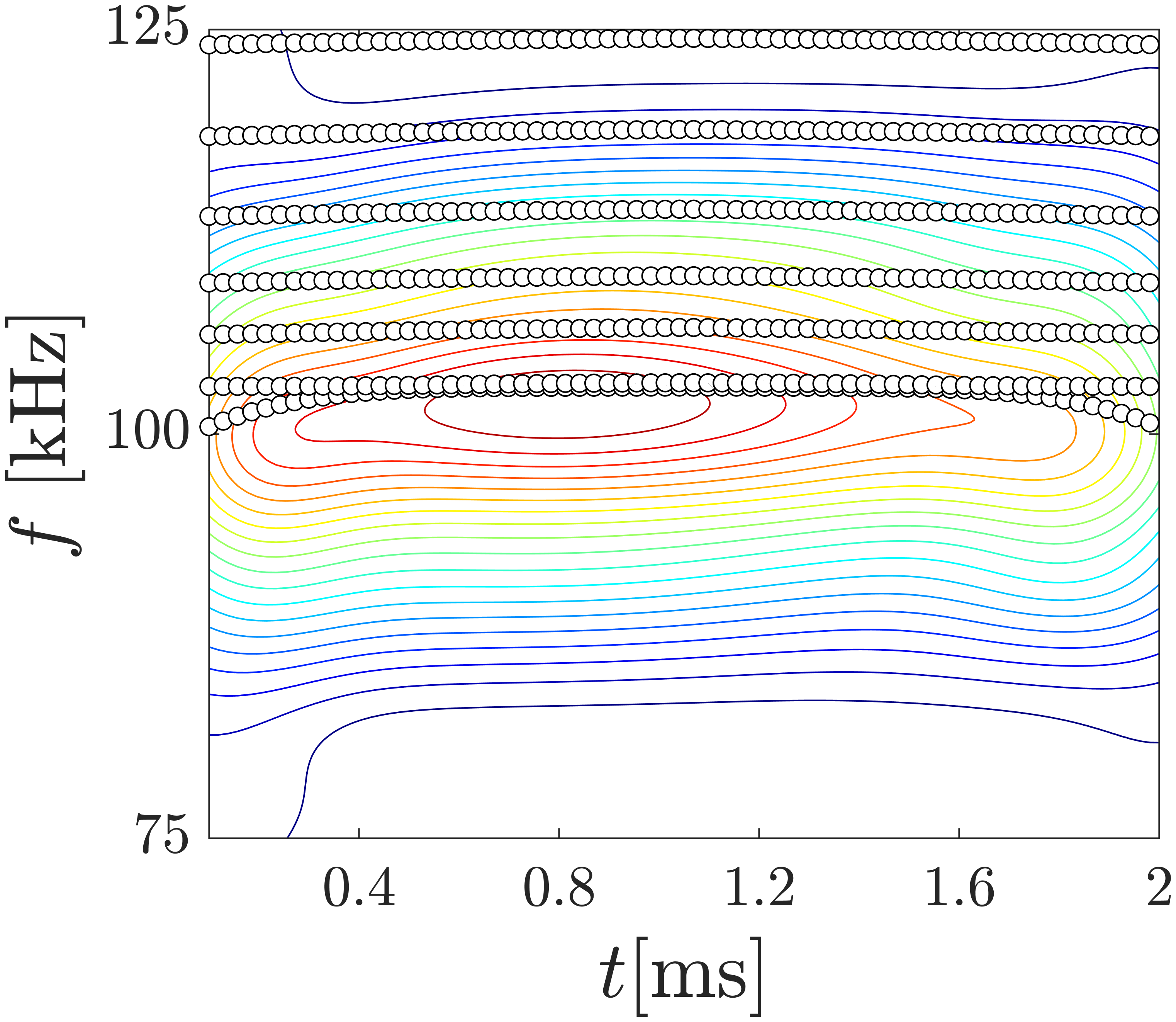}}
    \caption{Results for a temporally driven edge-to-edge mode conversion, which is achieved through a double array of modulated resonators. \subref{fig: snapshots upper & lower array} Three relevant snapshots before, during, and after time modulation. \subref{fig: upper & lower array - rms} RMS of the time history. \subref{fig: upper & lower array - spectrogram} Comparison between numerically reconstructed (colors) and expected (dots) evolution of the modes. There is a mode that migrates toward the bulk band, which is representative of a surface wave that transforms first into a bulk mode and later back-transforms into a surface wave.}
    \label{fig: Upper & Lowe Array_dyn_res}
\end{figure}

To further illustrate the functionalities of the metasurface, we now employ a smooth modulation that drives the energy transfer from the upper to the lower surface of a solid with finite dimensions along $y$. In here, a longer solid is considered, which is necessary to induce a slow transition between surface and bulk waves.
Results are reported in Fig. \ref{fig: snapshots upper & lower array} in terms of snapshots before, during, and after the modulation. 
To achieve this effect, a slow temporal drive is applied to the top and bottom layer $k_{r}=k_0\rightarrow 2.4\;k_0$, in a way that two coincident surface states belonging to the opposite edges form and, later, delocalize into a shear wave. In other words, due to an adiabatic mechanism, the surface wave slowly transforms into a shear wave. Then, once the shear mode is formed, the temporal drive $k_{r}=2.4\;k_0\rightarrow k_0$ is applied only to the bottom metasurface, thereby achieving localization in the form of a surface wave. 
The behavior is well summarized by the RMS of the entire time history reported in Fig. \ref{fig: upper & lower array - rms}, which further confirms the edge-to-edge energy transfer. The corresponding spectrogram with superimposed evolution of the states is displayed in Fig. \ref{fig: upper & lower array - spectrogram}. We observe that during the entire duration of the process, the energy, initially injected in the surface, follows the expected evolution of the surface states which, later, temporarily merge with the bulk. When the temporal drive is applied to the bottom surface, such a mode back-transforms into a surface wave traveling along the opposite surface, thereby achieving an edge-to-edge energy transfer.

\section{Conclusions}
This work illustrates the dynamics of time-varying metasurfaces. The implementation, which relies on time-modulated resonators, induces a wavenumber-preserving frequency transformation which is suitable to manipulate wave motion. The role of the modulation parameters in the frequency and mode conversion capabilities of time-varying metasurfaces are critically discussed and several functionalities inherently linked with the velocity and strength of the modulation are revealed. We have shown: pure frequency conversion, frequency conversion with energy scattering, concurrent frequency and mode conversion, and surface-to-surface energy transfer across the solid. The transition between these functionalities is delineated by the adiabatic theorem, which dictates a condition for the transformation to occur with or without scattering toward the wave modes populating the solid. 

We envision this new family of metasurfaces to suit the design of SAW devices with unusual transport capabilities, such as frequency conversion, mode conversion, and edge-to-edge transport.

\newpage

\includepdf[pages=-]{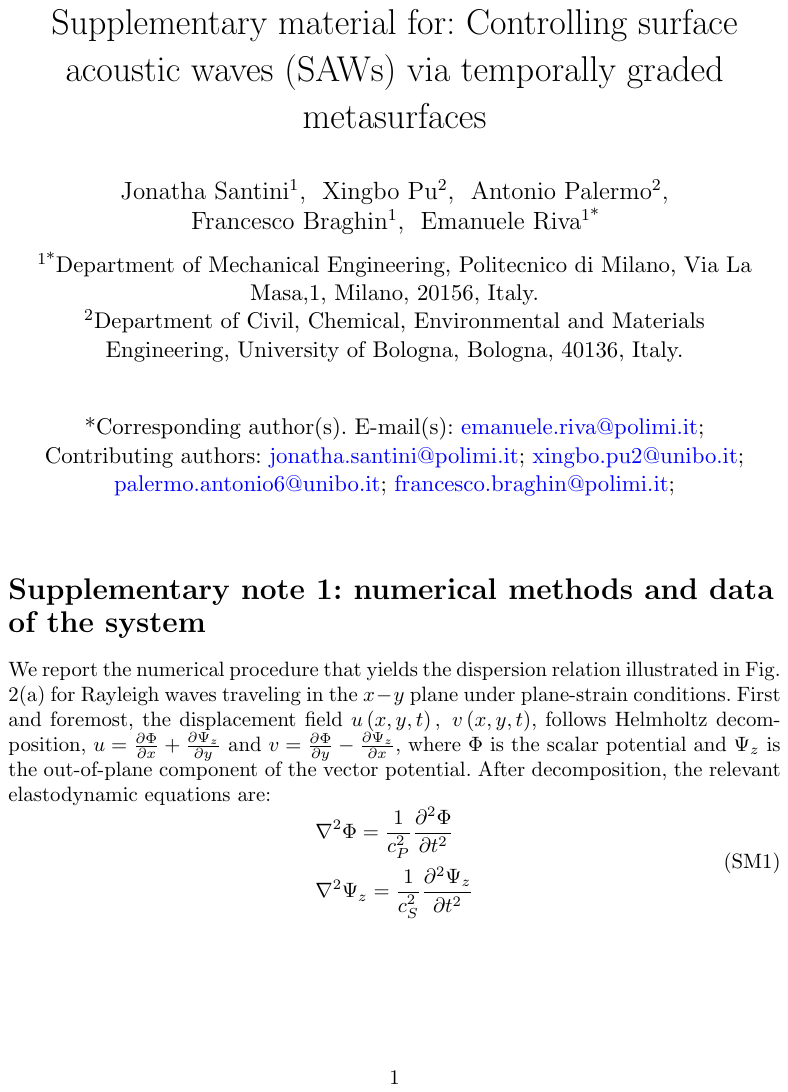}

\newpage

\bibliography{sn-bibliography}

\end{document}